\documentclass[aps,prl,twocolumn,superscriptaddress,twoside,floatfix,amsmath,showpacs]{revtex4}
\usepackage{dcolumn}
\usepackage{bm}
\usepackage[latin1]{inputenc}
\usepackage{graphicx}
\usepackage{amsmath}
\usepackage{amssymb}
\usepackage{amsfonts}
\usepackage{lineno}
\usepackage{url}
\usepackage{array}
\usepackage{verbatim}
\usepackage{subfigure}
\usepackage{color}
\usepackage{lscape}

\begin{document}

\title{Polarized Neutron Laue Diffraction on a Crystal Containing Dynamically Polarized Proton Spins}

\author{F.M.~Piegsa}
\altaffiliation{The principal authors, FMP and MK, contributed equally to this work. Electronic addresses: florian.piegsa@phys.ethz.ch (FMP); maths.karlsson@chalmers.se (MK).}
\affiliation{Institut Laue-Langevin, BP 156, F-38042 Grenoble, France}
\affiliation{ETH Z\"urich, Institute for Particle Physics, CH-8093 Z\"urich, Switzerland}

\author{M.~Karlsson}
\altaffiliation{The principal authors, FMP and MK, contributed equally to this work. Electronic addresses: florian.piegsa@phys.ethz.ch (FMP); maths.karlsson@chalmers.se (MK).}
\affiliation{European Spallation Source ESS AB, P.O. Box 176, SE-221 00, Lund, Sweden}
\affiliation{Department of Applied Physics, Chalmers University of Technology, SE-41296 G\"oteborg, Sweden} 

\author{B.~van den Brandt}
\affiliation{Paul Scherrer Institute, CH-5232 Villigen PSI, Switzerland}

\author{C.J.~Carlile}
\affiliation{European Spallation Source ESS AB, P.O. Box 176, SE-221 00, Lund, Sweden}

\author{E.M.~Forgan}
\affiliation{School of Physics and Astronomy, University of Birmingham, Birmingham B15 2TT, UK}

\author{P.~Hautle}
\affiliation{Paul Scherrer Institute, CH-5232 Villigen PSI, Switzerland}

\author{J.A.~Konter}
\affiliation{Paul Scherrer Institute, CH-5232 Villigen PSI, Switzerland}

\author{G.J.~McIntyre}
\affiliation{Institut Laue-Langevin, BP 156, F-38042 Grenoble, France}
\affiliation{Australian Nuclear Science and Technology Organisation, Locked Bag 2001, Kirrawee DC NSW 2232, Australia}

\author{O.~Zimmer}
\affiliation{Institut Laue-Langevin, BP 156, F-38042 Grenoble, France}

\date{\today}

\begin{abstract}


We report on a polarized-neutron Laue diffraction experiment on a single crystal of neodymium-doped lanthanum magnesium nitrate hydrate containing polarzed proton spins. By using dynamic nuclear polarization to polarize the proton spins, we demonstrate that the intensities of the Bragg peaks can be enhanced or diminished significantly, whilst the incoherent background, due to proton spin disorder, is reduced. It follows that the method offers unique possibilities to tune continuously the contrast of the Bragg reflections and thereby represents a new tool for increasing substantially the signal-to-noise ratio in neutron diffraction patterns of hydrogenous matter.

\end{abstract}


\maketitle

\begin{figure}
	\centering
		\includegraphics[width=0.45\textwidth]{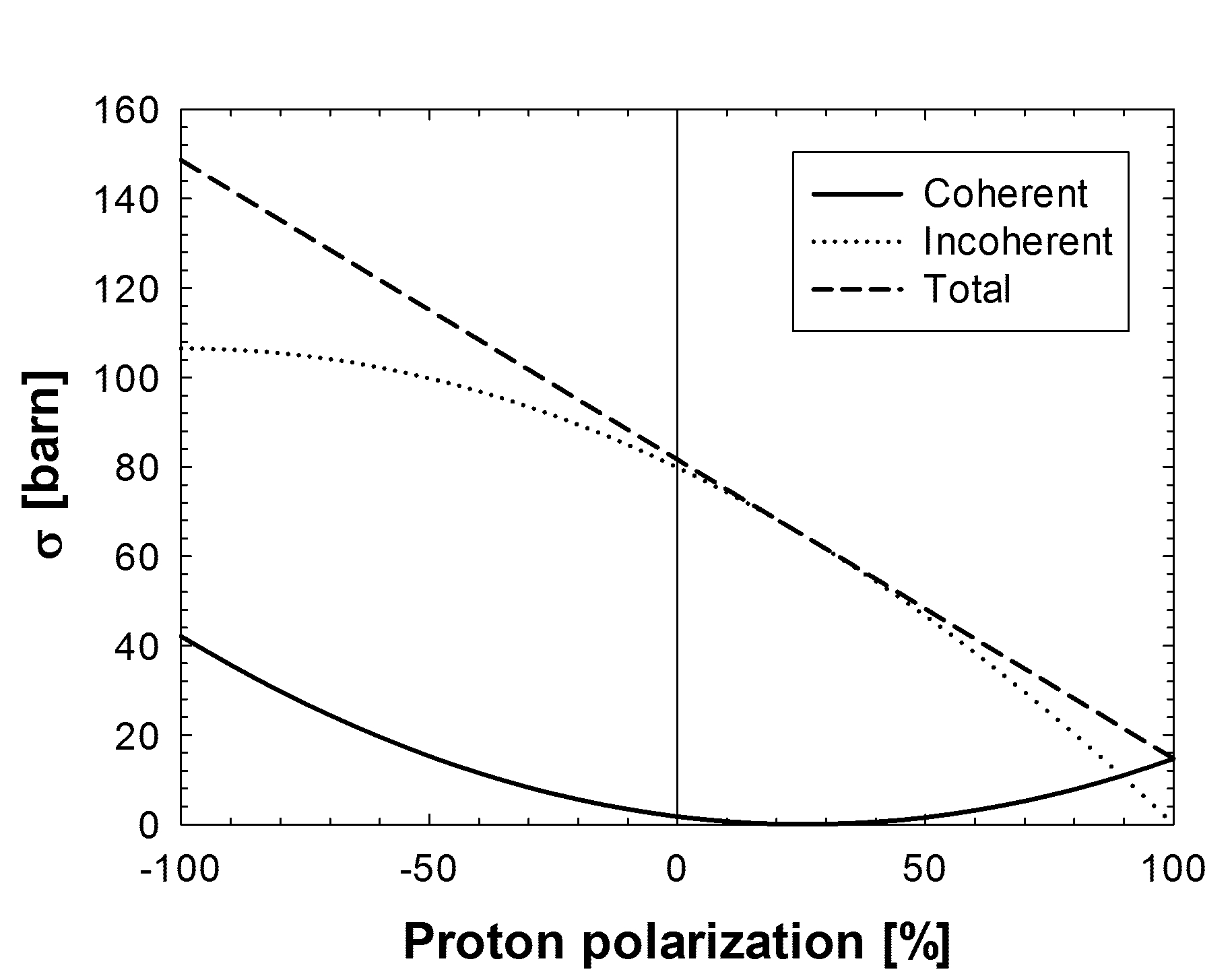}
	\caption{Coherent, incoherent and total scattering cross-section of hydrogen as a function of the proton polarization for fully polarized neutrons, i.e. $P_{\text{n}}=1$. }
	\label{fig:scatt}
\end{figure}

Hydrogen atoms play a key role in many materials and biological systems of high interest; examples are biomolecules, fuel cells and soft condensed matter.
Neutron diffraction studies of such hydrogenous matter generally suffer from a strong featureless background due to incoherent scattering by the protons. The incoherent scattering arises because the proton spins are normally completely disordered, and the scattering length - the strength of the neutron-proton interaction - depends strongly on the relative orientation of proton and neutron spins. 
A common way to reduce this incoherent scattering is to substitute the hydrogen by deuterium, since the incoherent scattering length of deuterium is far smaller than that of hydrogen. 
However, producing such samples is difficult, expensive and, for some materials (\textit{e.g.} complex biomolecules), often impossible. Equally well, isostructuralism cannot be guaranteed. 
Another method to reduce, or even remove completely, the incoherent scattering by the hydrogen atoms, is to align the spins of the neutrons and protons so that they are parallel \cite{ 30,50,3,51,52}. 
This is seen from Fig.~\ref{fig:scatt}, which shows the polarization-dependent coherent and incoherent cross sections for hydrogen, according to 
\begin{equation}
	\sigma_{\text{inc}} = 4\pi b_{\text{i}}^2 \left( 1 - \frac{P_{\text{n}} P_{\text{I}}}{I+1} - P_{\text{I}}^2\frac{I}{I+1} \right)~\text{and}
\end{equation}
\begin{equation}
	\sigma_{\text{coh}} = 4\pi  \left( b_{\text{c}}^2  + 2 P_{\text{n}} P_{\text{I}} b_{\text{c}} b_{\text{i}} \sqrt{\frac{I}{I+1}} + P_{\text{I}}^2 b_{\text{i}}^2 \frac{I}{I+1} \right)\text{,}
\end{equation}
where $P_{\text{I}}$ and $P_{\text{n}}$ are the polarizations of the protons and the incident neutron beam, respectively, $b_{\text{c}}$ is the
coherent scattering length of hydrogen, and $b_{\text{i}}$ is the incoherent \cite{30,13}. When the spins of the neutrons and the protons are parallel ($P_{\text{n}}P_{\text{I}}$ = 1) there is no incoherent scattering, and when they are anti-parallel ($P_{\text{n}}P_{\text{I}}$ = -1) the incoherent scattering is maximized. 
It is also evident from Fig.~\ref{fig:scatt} that the coherent scattering cross section increases from 1.8 to 14.7 barns when the spins are aligned, and hence a huge improvement in the signal-to-noise ratio will be observed.
The latter effect was originally demonstrated by Hayter et al. \cite{10} in the 1970's for a monochromatic neutron beam and a single crystal of neodymium-doped lanthanum magnesium nitrate hydrate, La$_2$Mg$_3$(NO$_3$)$_{12}$$\cdot$24H$_2$O (LMN:Nd).  
Here, we report on a polarized-neutron Laue diffraction experiment \cite{11} on a single crystal of LMN:Nd containing polarized protons, with the goal to demonstrate experimentally that a significant improvement of the signal-to-noise ratio of the Laue patterns can be achieved. 
The measurements were performed at the FUNSPIN beam line at the continuous spallation neutron source SINQ at the Paul Scherrer Institute, Switzerland \cite{6}.

A key aspect of the experiment described here is the method to achieve a sizable nuclear spin polarization of the protons in the single crystal sample. The \emph{brute force} approach, i.e. cooling a sample which is situated in a strong magnetic field to temperatures in the millikelvin range, will not lead to the desired result of a high nuclear polarization due to the excessively long nuclear spin relaxation times under these conditions \cite{40}. On the other hand, the so-called dynamic nuclear polarization (DNP) technique allows the polarization of nuclear spins in a solid under less restrictive conditions \cite{1,14}. As a prerequisite for DNP the sample needs to contain a certain amount of unpaired electrons (paramagnetic centers) - typically 0.1\% of the nuclear concentration. As the electron spins possess a much larger magnetic moment than the nuclear spins and a considerably shorter spin-lattice relaxation time compared to the nuclear spins, they easily reach polarizations of almost 100\% at moderate temperatures of 1 K and magnetic fields of 2.5 T. This electron spin ordering can be transferred to the nuclear spins by irradiation of microwaves close to the electron paramagnetic resonance frequency, due to the dipolar interaction between the nuclear and electron spins. LMN:Nd with its high content of hydrogen (proton density: $3.9 \times 10^{22}$~cm$^{-3}$) is a well-known crystal to be polarized by means of the \emph{solid effect} of DNP \cite{4,5,9,53,54} and hence represents a good candidate material to perform a proof-of-principle experiment. 

\begin{figure}
	\centering
		\includegraphics[width=0.495\textwidth]{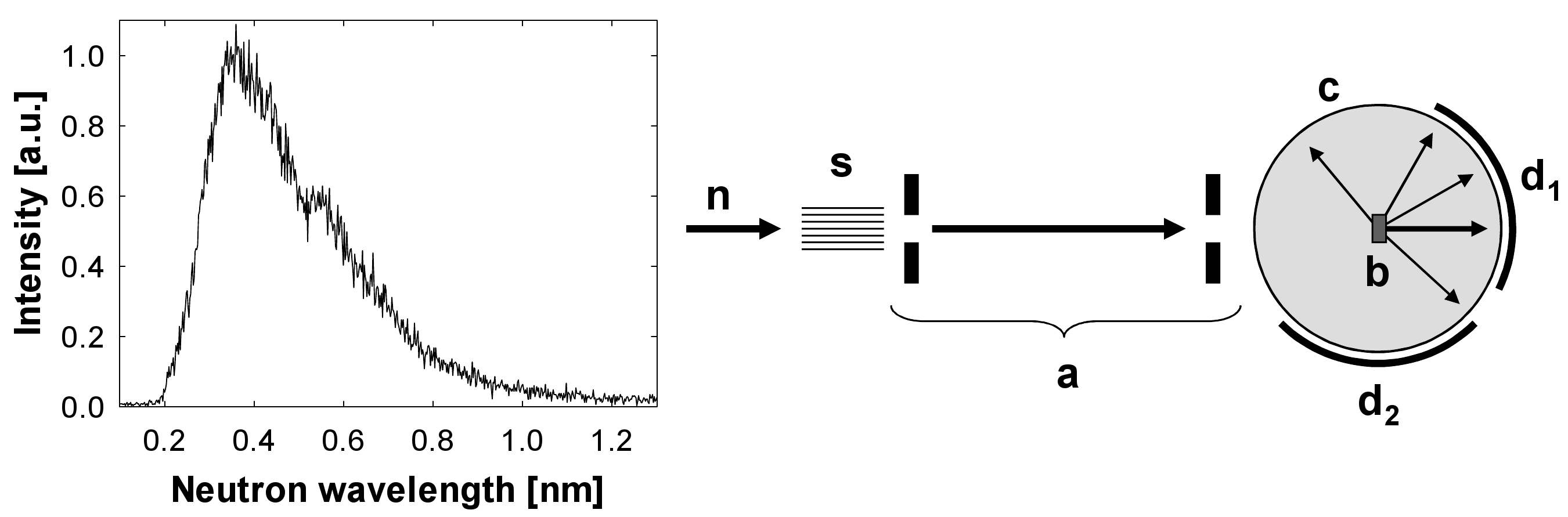}
	\caption{Left: spectrum of the polarized neutron beam with a peak flux at approximately 0.36 nm. This spectrum was determined by a separate time-of-flight measurement. Right: schematic top view of the polarized Laue setup. The incident neutron beam (n) is collimated by two Soller type collimators (s) and several diaphragms (a). The neutrons are scattered by the LMN:Nd crystal (b) situated in the DNP cryostat (c). The scattering pattern is recorded using standard neutron image plates (d$_1$ and d$_2$) which surround the cryostat.}
	\label{fig:setupTOF}
\end{figure}

A schematic drawing of the polarized neutron Laue setup is presented in Fig. \ref{fig:setupTOF} (right). The incoming white cold neutron beam (polarization $P_{\text{n}}\geq 95$\%) is collimated horizontally and vertically using an arrangement of two Soller collimators \cite {2} and several diaphragms, resulting in a circular beam with a diameter of 5 mm, a divergence of approximately 2 mrad (FWHM) and the spectral distribution presented in Fig. \ref{fig:setupTOF} (left). The neutron flux at the sample position is about $2.5 \times 10^5$ cm$^{-2}$s$^{-1}$ at a proton current of 1.5 mA on the SINQ spallation target.
The LMN:Nd crystal with a cross-section of $10 \times 10$ mm$^2$ and a thickness of 5 mm was placed inside a multi-mode microwave cavity which was top-loaded into a dedicated $^4$He evaporation DNP cryomagnet, with a base temperature of 1.05 K and a vertical split-pair 2.5 T magnet \cite{16}. Additionally, small permanent magnets were installed between the split pair, creating a weak horizontal holding field to avoid depolarization of the neutron beam as it passes through the zero-field region of the superconducting magnet. This horizontal field does not influence the homogeneity of the main magnetic field at the sample position, which is better than $2 \times 10^{-4}$. The latter is necessary to establish a uniform polarization over the entire sample. Positive proton spin polarization was achieved by irradiating with microwaves with a frequency of 78.965 GHz at a magnetic field of 2.06 T into the sample cavity using a 10 W Extended Interaction Oscillation Tube (electron $g$-factor in LMN:Nd: $g_{\bot}=2.70$). The proton polarization was measured with cw-NMR using a so-called Q-meter which was calibrated by means of the thermal equilibrium polarization \cite{15}.
In order to provide a constant nuclear polarization over several hours, continuous irradiation with microwaves was indispensable, as the nuclear spin relaxation time at 2.06 T and 1.05 K was determined to be about 150 min.

The Laue patterns were recorded using two standard neutron-sensitive image plates with a size of $400 \times 200$ mm$^2$ covering a total horizontal angle of about 210° and a vertical angle of $\pm 12$° (due to absorption of the scattered neutrons in the coils of the magnet only about half of the vertical dimension of the image plate could be used). The image plates were read out with a spatial resolution of 200 $\mu$m using a Fujifilm BAS-2500 scanner and the patterns obtained were normalized to the total neutron flux by means of a neutron monitor detector placed in the white beam. Furthermore, the neutron windows of the cryostat were made from aluminum with a thickness of only 0.2 mm, to reduce the background from neutrons scattered out of the incident and transmitted beams. 

\begin{figure}
	\centering
		\includegraphics[width=0.495\textwidth]{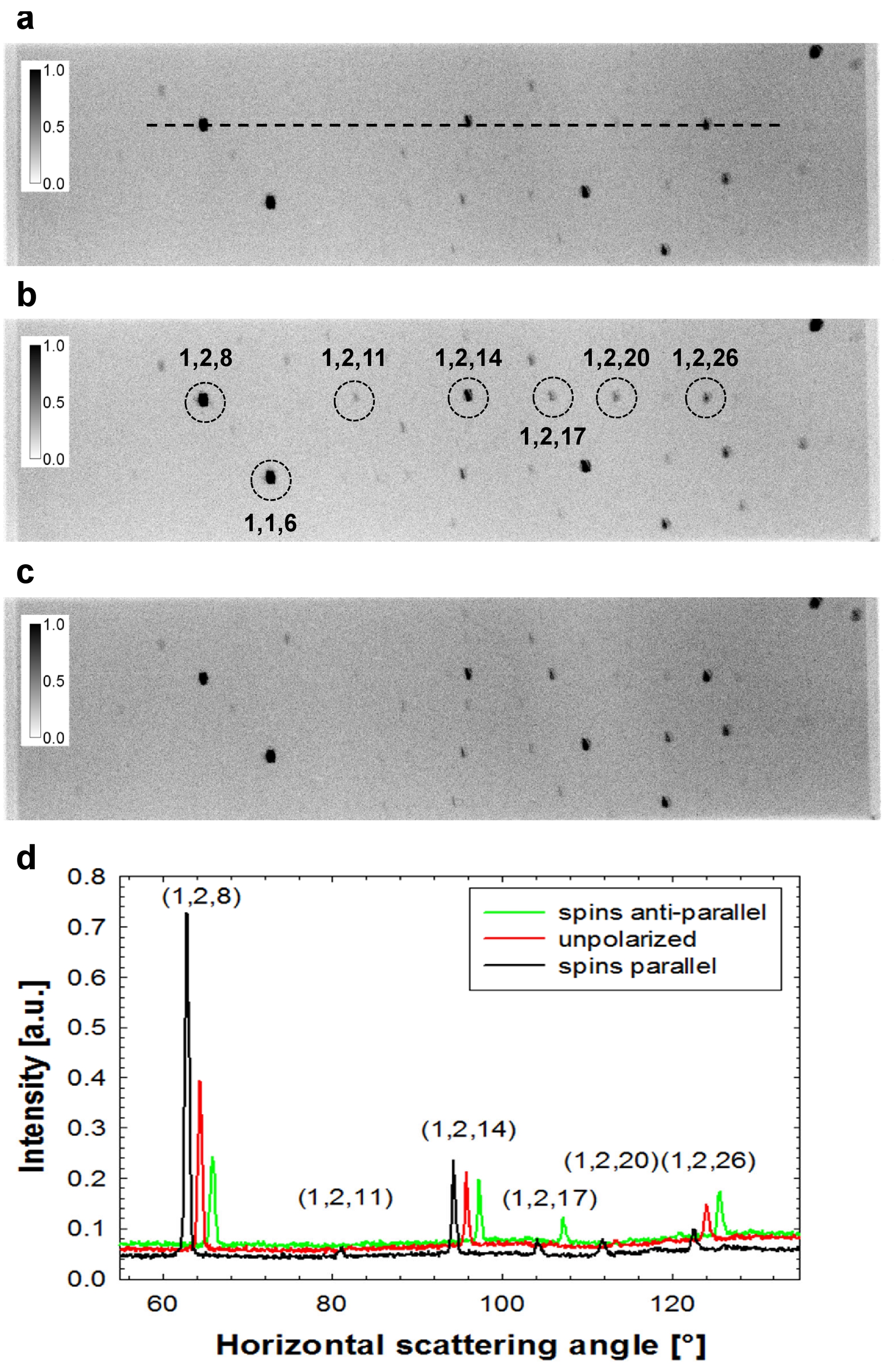}
	\caption{Normalized and background-subtracted Laue patterns of image plate d$_2$: a) unpolarized sample, b) proton and neutron spins parallel, and c) spins anti-parallel (proton polarization in both cases is approximately 35\%). d) Cuts through all three patterns along the dashed horizontal line indicated in a). For clarity the plots for parallel and anti-parallel spins have been horizontally shifted by -1.5° and +1.5° in horizontal scattering angle, respectively.}
	\label{fig:0080+0062+0065}
\end{figure}

In order to observe the effect on the neutron Laue pattern intensity and the incoherent background, we performed measurements with an unpolarized sample and with a sample with a proton polarization of 30-35\%. The polarization of the incoming neutron beam could be flipped by means of an adiabatic spin flipper, such that the spins of the protons and neutrons could be oriented either parallel or anti-parallel. The results for these three cases are depicted in Fig. \ref{fig:0080+0062+0065}. Each image shows the Laue pattern for the horizontal scattering angle from 41° to about 146°, which corresponds to image plate d$_2$ in Fig. \ref{fig:setupTOF} (right). The exposure time for each pattern was about 10 hours.
The instrumental background, i.e. the background from sources other than the incoherent scattering of the sample, is obtained in the same manner, but without the sample in the cryostat. This background contributes approximately 35\% to the total background in the case of the unpolarized sample and is subtracted from each of the three patterns.

The relative change of the incoherent scattering due to the proton polarization can be determined by averaging the neutron intensity over several areas of the pattern where there are no Laue reflections. In the case where the neutron and the proton spins are aligned parallel the incoherent scattering reduces to $(76 \pm 2)$\%, while it increases to $(115 \pm 2)$\% for anti-parallel spins. These results are in good agreement with the calculated values of 73-77\% and 117-119\%, if one assumes that the incoherent scattering is solely caused by the protons in the sample and a polarization of 30-35\%.

The Laue reflections were indexed using the program LAUEGEN \cite{20,21} and integrated using the program ARGONNE-BOXES \cite{22}, with the true counting statistics estimated from the observed point-by-point variations in the local background \cite{23}. The intensities for a selection of Laue reflections are summarized in Table \ref{tab:resultstab}. It is seen that the intensities  of the individual reflections can be changed significantly by spin polarization. For example, in an unpolarized sample, the reflection (1,2,11) is practically undetectable, while it becomes clearly visible in the case of parallel spins. Significant changes in intensity with polarization are also observed in some reflections in the row $(1,1,l)$. 
Because the intensity of Bragg reflections may be thereby increased or decreased, an improvement of the signal-to-noise ratio is not observed for all the recorded reflections.
However, we do note that for large-unit-cell biological crystals, the intensity of incoherent background compared to that of the Laue reflections will be much greater than for our proof-of-principle experiment. Hence, the reduction of this background by polarization of the sample will be even more important in that case. 
\begin{table*}
	\centering
		\begin{tabular}{|>{\centering\arraybackslash} p{1.2cm}|>{\centering\arraybackslash} p{1.2cm}|>{\centering\arraybackslash} p{1.2cm}|>{\centering\arraybackslash} p{2cm}|>{\centering\arraybackslash} p{2cm}|>{\centering\arraybackslash} p{2cm}|>{\centering\arraybackslash} p{2cm}|>{\centering\arraybackslash} p{2cm}|>{\centering\arraybackslash} p{2cm}|}
		\hline
		$h$ $k$ $l$  &  mult.  &  $\lambda$ $\left[ \text{nm} \right]$    &  $I_{\uparrow-}$  &  $I_{\uparrow\uparrow}$  &    $I_{\uparrow\downarrow}$ &    $I_{\uparrow\uparrow}/I_{\uparrow-}$ &    $I_{\uparrow\downarrow}/I_{\uparrow-}$ &    $I_{\uparrow\uparrow}/I_{\uparrow\downarrow}$    \\
		\hline
		\hline
		1 1 6         &    3    &   0.4986               & $114290 (512)$  & $136192 (520)$  & $76128 (448)$  & 1.19 & 0.67 & 1.79      \\
		1 2 8         &    1    &   0.3154               & $102135 (488)$  & $204047 (616)$   & $41663 (368)$ & 2.00 & 0.41 & 4.90      \\
		1 2 11        &    1    &   0.3281               & $108 (288)$     & $4513 (272)$    & $756 (248)$    & 41.8 & 7.00 & 5.97      \\
		1 2 14        &    1    &   0.3144               & $35023 (376)$   & $48455 (384)$   & $25683 (328)$  & 1.38 & 0.73 & 1.89      \\
		1 2 17        &    1    &   0.2920               & $974 (296)$     & $9020 (296)$    & $10473 (304)$  & 9.26 & 10.8 & 0.86      \\
		1 2 20        &    1    &   0.2682               & $3266 (304)$    & $8597 (304)$    & $1299 (280)$   & 2.63 & 0.40 & 6.62      \\
	  1 2 26        &    1    &   0.2257               & $21228 (360)$   & $12141 (320)$   & $22882 (352)$  & 0.57 & 1.08 & 0.53      \\
		\hline
		\end{tabular}
	\caption{Iintensities and intensity ratios for a selection of Laue reflections obtained from the data presented in Fig. \ref{fig:0080+0062+0065} for the sample unpolarized ($I_{\uparrow-}$), polarized parallel ($I_{\uparrow\uparrow}$) and anti-parallel ($I_{\uparrow\downarrow}$) with respect to the neutron spin, respectively.}
	\label{tab:resultstab}
\end{table*}

In conclusion, we have shown that the polarization of protons opens up the possibility to tune the contrast of the Bragg reflections, whilst the incoherent scattering is reduced, in a quite predictable manner. This concept of utilizing the strong spin dependence of the scattering cross sections of hydrogen may be very favorably employed to improve substantially the poor signal-to-noise ratio in neutron diffraction experiments on samples with a large hydrogen content. It follows that the technique may provide answers to specific questions on \textit{e.g.} hydrogen positions or protonation states and hence may have a huge impact on a large range of disciplines, ranging from hard and soft condensed matter to pharmaceutical biology. For the wide-spread use of dynamic nuclear polarization in neutron Laue diffraction, the method is however accompanied by a few demanding challenges and open questions:
(i) Which paramagnetic centers can be used and can they be implanted in the crystal, notably a biological crystal, of a sufficient size?
(ii) What is the level of proton polarization that can be achieved? 
(iii) What is the gain compared to deuteration?
(iv) How can the sample environment be optimized for Laue diffraction?
Although the answering of these questions requires further exploration, we can here give tentative responses:

\begin{itemize}
\item[(i')]
DNP on biological macromolecules was originally performed by adding a small percentage of a bulky Cr$^{5+}$ complex, 
C$_{12}$H$_{22}$CrO$_{4}$Na.H$_{2}$O to a solution containing the macromolecule of interest, which was then frozen to form a glassy sample \cite{30,80}. The paramagnetic carrier would generally need to be considerably smaller than this complex to promote or maintain translational symmetry in the crystalline state. The smaller TEMPO free radical, recently also used to create \emph{hyperpolarized} substances for biological MRI \cite{31}, would be the first choice.
\item[(ii')]
The polarization of 35\% achieved in this experiment is particularly unsuitable for determination of hydrogen positions since it corresponds to a coherent cross section of hydrogen of nearly zero (Fig. 1).  Achieving a polarization of at least 50\% is desirable and seems possible with state-of-the-art DNP equipment. However, it might be necessary to reduce the sample temperature further and/or increase the magnetic  field strength to improve the nuclear spin relaxation time, which constrains the maximum polarization. Much of the further development to increase the polarization can be done however without neutrons, using, e.g. NMR, to measure the polarization.
\item[(iii')] 
Where it is possible and does not change the structure or function, deuteration is at present still preferable to proton polarization, since a given percentage average deuteration can yield the same reduction in incoherent background as the same percentage proton polarization. However, when deuteration is not possible or for structure determination, the polarization technique offers unique possibilities.
\item[(iv')]
The intrinsically smaller sample sizes available and desirable for Laue diffraction, reduces the requirement on the spatial extend of the field homogeneity and the gap between the magnet pole pieces which should allow a scaling down of the appartus and/or an enlargement of the solid angle of detection.
\end{itemize}

We gratefully acknowledge help by the neutron radiography group of PSI (C. Gr\"unzweig, E.H. Lehmann and P. Vontobel), J. Kohlbrecher for fruitful discussions and P. Schurter for excellent technical support. 
Special microwave equipment was kindly provided to PSI by W.Th. Wenckebach.  M.K. and C.C. thank the Swedish Research Council for instrument development funding.\\ 
This work was performed at the Swiss Spallation Neutron Source at the Paul Scherrer Institute, Villigen, Switzerland.

\appendix

\end{document}